\documentclass{PoS}
\usepackage{amsmath,amstext}
\usepackage{wrapfig}
\usepackage[rawfloats=true]{floatrow}

\title{Modeling the Aperture of Radio Instruments for Air-Shower Detection}

\ShortTitle{Aperture of Radio Instruments}

\author{\speaker{V.~Lenok}$^{a}$,
P.~Bezyazeekov$^{b}$,
N.~Budnev$^{b}$,
O.~Fedorov$^{b}$,
O.~Gress$^{b}$,
O.~Grishin$^{b}$,
A.~Haungs$^{a}$,
T.~Huege$^{a,c}$,
Y.~Kazarina$^{b}$,
M.~Kleifges$^{d}$,
E.~Korosteleva$^{e}$,
D.~Kostunin$^{f}$
L.~Kuzmichev$^{e}$,
N.~Lubsandorzhiev$^{e}$,
S.~Malakhov$^{b}$,
T.~Marshalkina$^{b}$,
R.~Monkhoev$^{b}$,
E.~Osipova$^{e}$,
A.~Pakhorukov$^{b}$,
L.~Pankov$^{b}$,
V.~Prosin$^{e}$,
F.~G.~Schr\"oder$^{a,g}$
D.~Shipilov$^{b}$,
A.~Zagorodnikov$^{b}$
--- the Tunka-Rex Collaboration
~\\
\llap{$^{a}$}Institut f\"ur Kernphysik, Karlsruhe Institute f\"ur Technology (KIT), Karlsruhe, 76021 Germany\\
\llap{$^{b}$}Institute of Applied Physics ISU, Irkutsk, 664020 Russia\\
\llap{$^{c}$}Astrophysical Institute, Vrije Universiteit Brussel, Pleinlaan 2, 1050 Brussels, Belgium\\
\llap{$^{d}$}Institut f\"ur Prozessdatenverarbeitung und Elektronik, Karlsruhe Institute of Technology (KIT), Karlsruhe, 76021 Germany\\
\llap{$^{e}$}Skobeltsyn Institute of Nuclear Physics MSU, Moscow, 119991 Russia\\
\llap{$^{f}$}DESY, Zeuthen, 15738 Germany\\
\llap{$^{g}$}Bartol Research Institute, Department of Physics and Astronomy, University of Delaware, Newark, DE, 19716, USA\\
E-mail: \email{vladimir.lenok@kit.edu}
}
    
\abstract{
Sparse digital antenna arrays constitute a promising detection technique for future large-scale cosmic-ray observatories.
It has recently been shown that this kind of instrumentation can provide a resolution of the energy and of the shower maximum on
the level of other cosmic-ray detection methods.
Due to the dominant geomagnetic nature of the air-shower radio emission in the traditional frequency band of 30 to 80 MHz,
the amplitude and polarization of the radio signal strongly depend on the azimuth and zenith angle of the arrival direction.
Thus, the estimation of the efficiency and subsequently of the aperture of an antenna array is more complex
than for particle or Cherenkov-light detectors.
We have built a new efficiency model based on utilizing a lateral distribution function as a shower model, and
a probabilistic treatment of the detection process.
The model is compared to the data measured by the Tunka Radio Extension (Tunka-Rex),
a digital antenna array with an area of about 1\,km$^2$ located in Siberia at the Tunka Advanced Instrument for Cosmic rays and Gamma Ray Astronomy (TAIGA).
Tunka-Rex detects radio emission of air showers using trigger from air-Cherenkov and particle 
detectors.
The present study is an essential step towards the measurement of the cosmic-ray flux with Tunka-Rex,
and is important for radio measurements of air showers in general.
}

\FullConference{36th International Cosmic Ray Conference -ICRC2019-\\
		July 24th - August 1st, 2019\\
		Madison, WI, U.S.A.}

\begin{document}

\section{Introduction}
The estimation of efficiency and aperture of the radio detectors for cosmic-ray air showers is currently one of the unsolved problems
in observational ground-based astroparticle physics.
Without proper understanding of this quantity we cannot correctly address determination of
the energy spectrum
and mass composition from radio measurements.
In this work we present a new approach for the estimation of the efficiency of a radio instrument
based on a statistical treatment of the detection process and using a model of lateral distribution 
of 
the radio signals.

The model was developed for applications to the Tunka-Rex instrument, and is under validation 
against the hybrid Tunka-Rex/Tunka-133 data.

\textbf{Tunka-Rex instrument}.
Tunka Radio Extension (Tunka-Rex) is a digital radio antenna array for observations of cosmic-ray 
showers via their radio emission~\cite{BEZYAZEEKOV201589}.
The array is located on the site of the TAIGA detector at the
Tunka valley in Siberia~\cite{Kuzmichev:2018mjq}.
The instrument has evolved over time and for the time being comprises of 63 antennas (about 1\,km$^2$ area) of
short aperiodic loaded loop antenna (SALLA) type~\cite{Abreu_2012} operating in the traditional frequency range of 30--80\,MHz.
57 of those antennas are in the core region of the instrument, and 6 antennas are in the satellite clusters.
The instrument operates since 2012.

Tunka-Rex receives trigger from the two co-located detectors of the TAIGA observatory --- 
the Cherenkov-light detector Tunka-133 and the particle detector named Tunka-Grande.
However, for the present analysis we use only Tunka-133 triggered data.

\section{Available Data}

For the present analysis we used all measurements of Tunka-Rex jointly with Tunka-133
(Tunka-Grande triggered data is not included) collected during 2012--2017.
The reason for this will be discussed below.
The observation time per season with the corresponding antenna configuration is shown in Tab.~\ref{time}.
The configurations are divided into so-called generations --- according to the construction stage of the array.

\begin{table}[h!]
\begin{center}
\begin{tabular}{ l l l l }
\hline
\hline
Gen. & Number of antennas & Operation seasons & Operation time \\
\hline
1a & 18 & 2012/13 & 282.686(83)\,hours \\
1b & 25 & 2013/14 & 214.467(83)\,hours \\
2 & 44 & 2015/16 & 305.767(83)\,hours \\
3 & 63 & 2016/17 & 265.550(83)\,hours \\
\hline
  &    & Total   & 1068.4667(83)\,hours \\
\hline \hline
\end{tabular}
\caption{Operation time of the Tunka-Rex detector in all seasons (only Tunka-133 triggered).}
\label{time}
\end{center}
\end{table}

\begin{figure}
  \includegraphics[width=\linewidth]{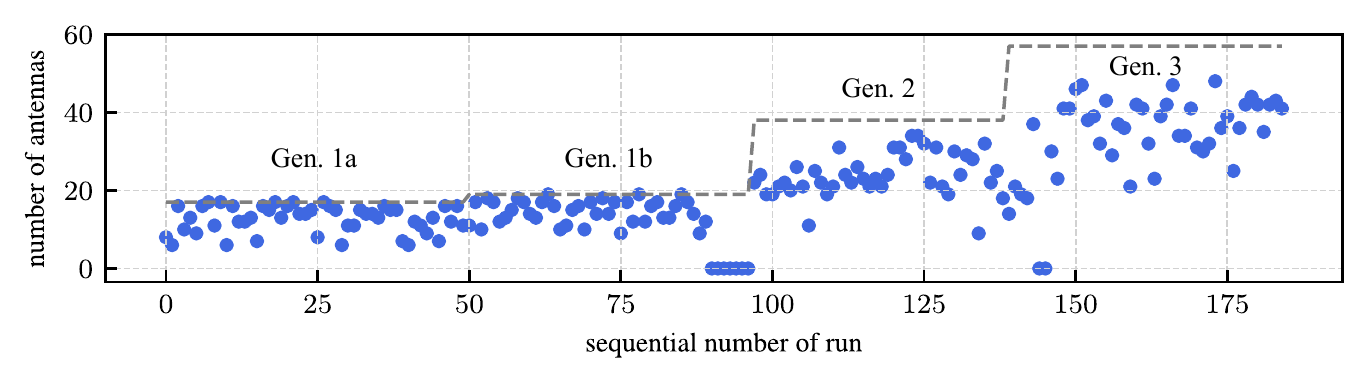}
  \vspace{-10mm}
  \caption{The number of operating antennas in each considered runs (one run per night with 
Tunka-133 operation).
  The dashed line above indicates the
  number of the installed antennas for a given run.}
  \label{up}
\end{figure}

However, not all installed antennas
are available for measurements all the time due to
electronic and other kinds of malfunctions.
This makes the actual configuration of the array different for each of the observational runs.
To obtain this information we performed an analysis of all radio events
(including those that do not have 3-fold coincidence --- the standard condition for an event).
The number of operating antennas for each of the investigated runs is shown in Fig.~\ref{up}.
This information is important for a correct estimation of the efficiency of the instrument.

\begin{wrapfigure}{r}{0.5\textwidth}
\vspace{-5mm}
  \begin{center}
    \includegraphics[width=1.\linewidth]{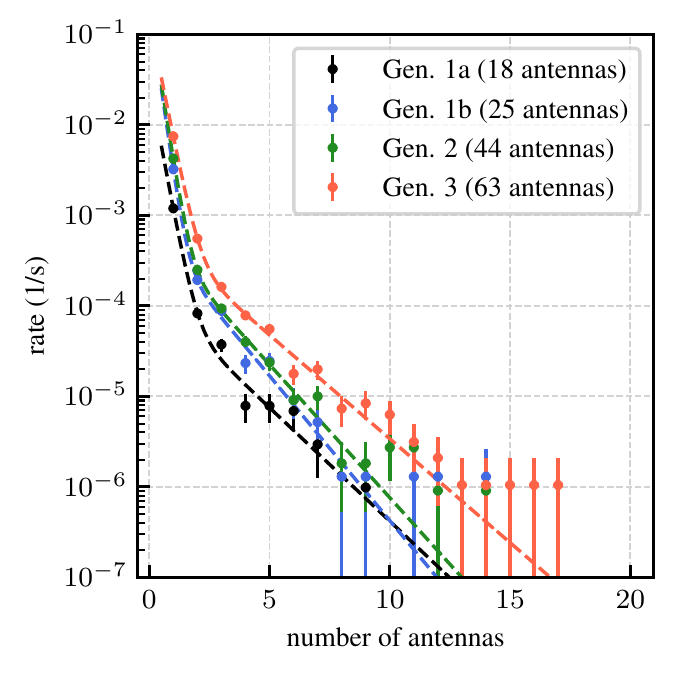}
  \end{center}
  \vspace{-10mm}
  \caption{Multiplicity of the detected events.}
  \label{mult}
    \vspace{-10mm}
\end{wrapfigure}
Additional information on the measurements can be obtained from the event multiplicity plot (Fig.~\ref{mult}).
For each of the data set you can see auxiliary lines of a sum of two exponential laws fitted to the 
data.
It is clearly seen that the events with multiplicity less then three 
have a higher rate, probably due to background.
This supports the standard requirement of a 3-fold coincidence for cosmic-ray air-shower events.

\section{Aperture Estimation}
The radio emission from a cosmic-ray air shower observed in the traditional frequency range
depends on the incoming direction with respect to the geomagnetic field.
Thus, the instrument efficiency $\epsilon$ of radio detectors operating in the traditional frequency range 
is a function of the incoming direction (zenith $\theta$ and azimuth $\phi$ angles), the shower core 
position $\vec{r}$, the depth of the shower maximum $X_{max}$, and the energy of the primary 
particle.
The efficiency is the ground of the aperture.
That complexity of the efficiency behavior makes the estimation of the aperture for radio 
instruments difficult.

The definition of the aperture $A$ is
\begin{equation}
 A = \int\limits _{\Omega}  \iint \limits _{S'} \epsilon \, d \vec{r'} \, d \omega =
 \int\limits_{\Omega} \iint\limits _{S} \epsilon \cos \theta d \vec{r} \, d \omega.
\end{equation}

The initial definition is the integration of the efficiency $\epsilon$ over the
viewing solid angle of the instrument $\Omega$ and the
projection of its fiducial area $S'$ to the plane perpendicular to the incoming direction.
For modeling a flat horizontal detector we introduce a cosine factor depending
on the zenith angle, and then we can integrate over the instrument fiducial area $S$
(without the projection).

There are two problems in the aperture estimation:
estimation of the shape of the efficiency function itself, and
calculation of the integral with sufficient precision.
Estimation of the integral can introduce additional uncertainties on the level of several percent,
and requires using the special methods~\cite{Beentjes2016QUADRATUREOA}.
Also, it was shown recently that there is a way to evaluate the integral semi-analytically
for the full efficiency regions~\cite{Lenok:2018das}.
Evaluation of these methods in application is subject for
further studies.
In the present work we address the former problem --- estimation of the shape of efficiency function.

\section{Efficiency Model}

The widely used method of the efficiency estimation by Monte Carlo
simulation of the instrument performance is difficult to apply to the radio arrays
due to high computational expenses for calculating the radio emission from a single shower.

\textbf{General approach}.
In order to avoid the generation of a large simulation library
we use models based on
lateral distribution functions of the radio emission on the detection plane.
Namely, we reverse the lateral distribution functions used in the shower reconstruction
to the prediction of the signals from the given shower parameters, to evalute the efficiency.

First, we will briefly recall the essentials of the existing Tunka-Rex efficiency model.
Then we will present a new developed model of efficiency.

\textbf{Footprint-based Model}.
In a simplified approach a model based on a parametrization of the shower-footprint
size can be considered.
This type of model was developed by Tunka-Rex, and was used for general 
purposes~\cite{Fedorov:2017xih, PhysRevD.97.122004}.
The estimation of the radio footprint size was done
using the exponential LDF, which was initially used in the data analysis,
and the detection-threshold estimation based on the distribution of the background amplitudes.
Calculation of the efficiency in this model is done by counting antennas within the footprint.
Due to the sharply defined border of the radio footprint the model has a pronounced binary nature
--- a shower is either detected or not.

\begin{figure}[h!]
\center{
  \includegraphics[width=0.9\linewidth]{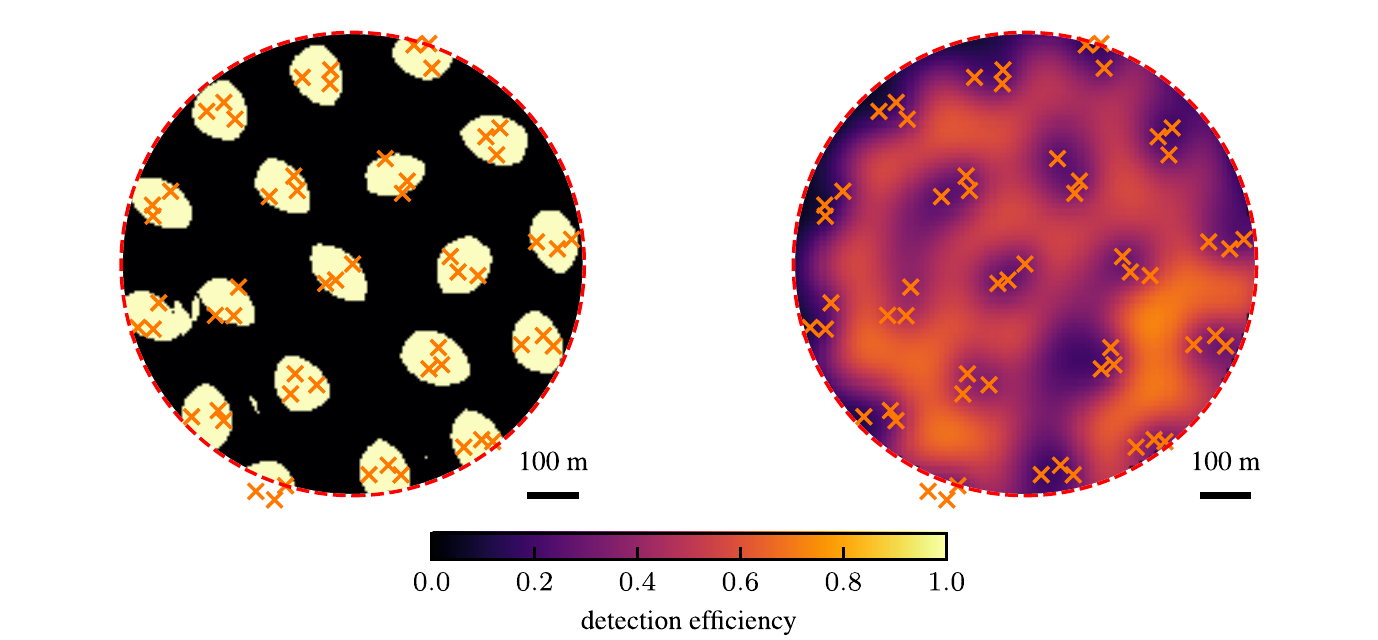}
  }
  \vspace{-0mm}
  \caption{Performance of the efficiency models
          ($\vartheta$\,=\,30$^{\circ}$, $\varphi$\,=\,60$^\circ$, $\lg$(E/eV)\,=\,17.1).
          The outer antennas are not shown.
          \textit{Left}: \textbf{the footprint-based model} that uses
          the footprint border parametrization.
          \textit{Right}: \textbf{the LDF-based probabilistic model} that takes into account the
          antenna detection probabilities. (Dashed circles --- the fiducial area of Tunka-Rex.)}
  \label{efficiency}
\end{figure}

The example of the model performance is shown in Fig.~\ref{efficiency}.
The applications of this model to derive of the energy spectrum can be found in 
Ref.~\cite{Kostunin_ICRC2019}.

\textbf{LDF-based Model}.
This new model uses 
the LDF used in the Tunka-Rex data analysis,
the single antenna detection threshold,
and the probabilistic treatment of the event detection.
All of these components makes this model more realistic.

\begin{wrapfigure}{r}{0.5\textwidth}
\vspace{-5mm}
  \begin{center}
    \includegraphics[width=1.\linewidth]{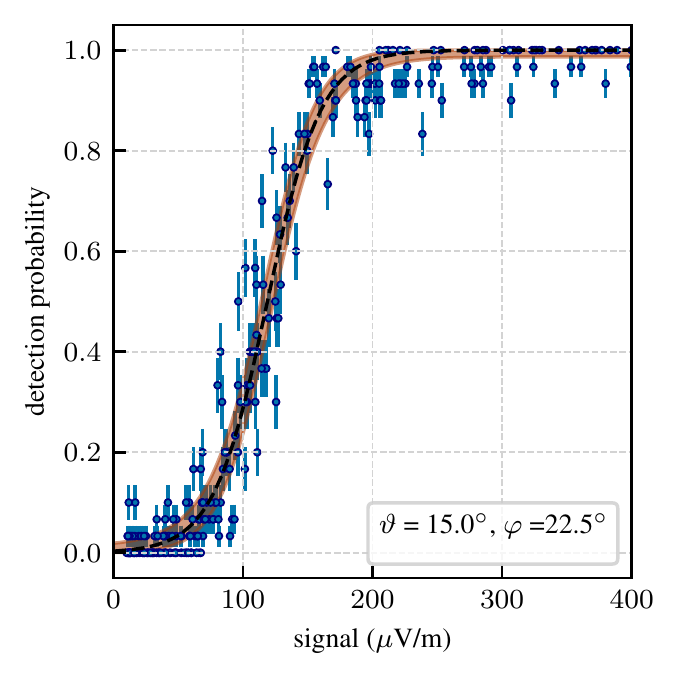}
  \end{center}
  \vspace{-5mm}
  \caption{The single antenna detection probability as function of the signal amplitude.
	  The points are results of the simulation analysis.
	  The curve is the logistic function used as a model.
	  The angles are the position of the bin center.}
    \label{threshold}
    \vspace{-10mm}
\end{wrapfigure}

The radio antenna holds a probabilistic nature of the signal detection due to
the presence of different types of noise
and its fluctuations in time, which is particularly pronounced in case of small signals.
For very large signals, when the signal-to-noise ratio is high, this effect is negligible.
To examine this behavior we studied the detection probability 
with simulations of the signal and measured noise.
The signals from the simulations with randomly distributed incoming directions were processed with
the Tunka-Rex analysis pipeline excluding the stages related to
the reconstruction of the shower parameters.
Since the probabilistic behavior is caused by the noise in traces,
each simulated signal was reprocessed 30 times with different noise samples.
The Tunka-Rex standard noise library was used for that purpose.
The probability was defined as a number of times when the signal with a given noise passed
the selection criteria of the Tunka-Rex signal processing procedure.
The data were binned by angles (bin widths: $\Delta\theta$\,=\,10$^{\circ}$, $\Delta\phi$\,=\,45$^{\circ}$)
and further analyzed bin-wisely to have sufficient coverage of amplitudes.
In Fig.~\ref{threshold} one can see the results of the analysis for one of the bins.
As a model of detection probability we chose the logistic function:
\begin{equation}
 p = \left( 1 + \exp \left[-k \left(S-S_{1/2} \right) \right] \right)^{-1}.
\end{equation}
$S$ is a signal amplitude, $S_{1/2}$ is a signal with 1/2 probability to be detected,
$k$ is a parameter.

In order to achieve a homogeneous coverage over the whole range of considered angles,
we introduce a simple linear model of the logistic curve parameters as function of
the zenith angle.
Namely
\begin{equation}
        k = k' + k'' \vartheta, \hspace{5mm}
        S_{1/2} = S_{1/2}' + S_{1/2}'' \vartheta.
\end{equation}
The parameters of the model were fit to the simulation data and are as follows
\begin{equation}
 \begin{aligned}
        k'    &= 5.580(134)  \times 10 ^{-2} \, m / \mu V, \\
        k''   &= -0.912(265) \times 10 ^{-2} \, m / \mu V, \\
        S_{1/2}'  &= 1.167(147) \times 10 ^{2} \, \mu V / m, \\
        S_{1/2}'' &= 1.131(299) \times 10 ^{1} \, \mu V / m.
  \end{aligned}
\end{equation}
Despite the fact that the uncertainties of the parameters are significant, this approach
allows us to use a direction-dependent antenna threshold which brings the analysis closer to reality.
The parametrization can be improved in future by including more simulation data and thus applying
a finer binning, non-linear terms, etc.

\begin{wrapfigure}{r}{0.5\textwidth}
\vspace{-5mm}
  \begin{center}
    \includegraphics[width=1.\linewidth]{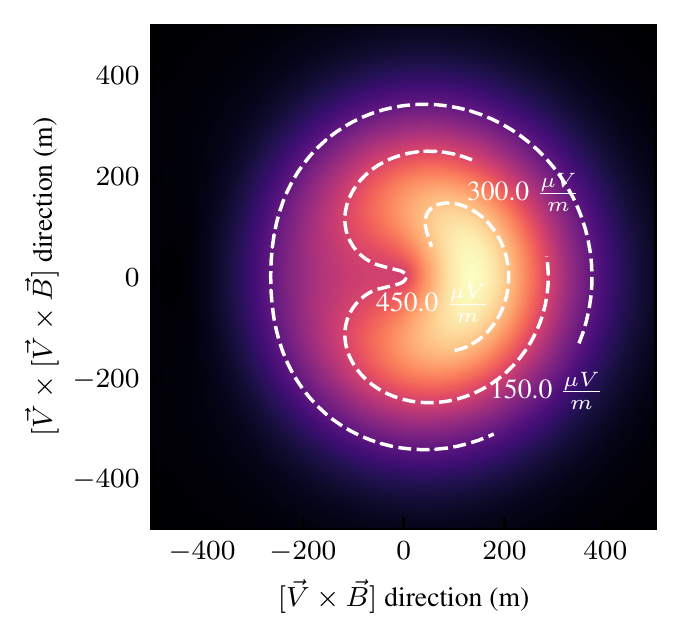}
  \end{center}
  \vspace{-5mm}
  \caption{An example of the Tunka-Rex LDF.
	    The distribution shows the strength of the shower electric field corresponding to the signal 
maxima
	    with respect to the geomagnetic coordinate system.
            E~=~1~EeV, X$_{max}$~=~600~g/cm$^2$, $\vartheta$~=~40$^{\circ}$, 
$\alpha$~=~20$^{\circ}$.}
    \label{trex_ldf}
    \vspace{-5mm}
\end{wrapfigure}

The model which is used in the standard Tunka-Rex reconstruction is now reversed into
the prediction of the expected radio signals from given parameters of a shower.
The model of spatial distribution of the signals $S$ from the reversed Tunka-Rex analysis is
as follows (Tunka-Rex LDF model):
\begin{equation}
\begin{aligned}
        S   &= S_0 \left[ \xi^2(r) + 2 \xi(r) \cos \varphi_t \sin \alpha + \sin ^2 \alpha 
\right]^{-1/2 },\\
        S_0 &= S_{r_0} \exp \left[ -a (r - r_0)^2 + b(r - r_0) \right], \\
        S_{r_0} &= \frac{E}{\varkappa} \, \exp \left[ a(x' - x_0)^2 - b(x' - x_0) \right], \\
        \xi &= \xi_0 r + \xi_1 r^2 + \xi_2 r^3, \\
        a &= \left( a_0 + a_1 E \right) + \left( a_2 + a_3 E \right) \cos \vartheta, \\
        b &= b_0 - \exp \left[ \left( X_{0} / \cos \vartheta - X_{max} - b_1 \right)/b_2 
\right].
 \end{aligned}
\end{equation}
$r$ is a distance to the shower axis,
$X_{0}$ is a vertical depth of detection level (955\,g/cm$^2$ for Tunka-Rex),
$\alpha$ is a geomagnetic angle,
$\varphi_t$ is a geomagnetic azimuth,
$x'$ and $x_0$ are reference distances,
$\xi_i$, $a_i$, $b_i$ are constants.
An example of the LDF obtained with this model can be seen in Fig.~\ref{trex_ldf}.

We apply the
earlier described detection probability model for a single antenna to the
signals predicted by this LDF model.

To obtain the probability to detect an air shower from a set of the single probabilities 
we make the following calculation.
Since the shower is considered as detected 
when $n$ and more antennas have signals out of all $N$ antennas,
we sum the probabilities of all realizations that are not leading to the trigger condition
and then subtract it from unity:
\begin{equation}
 P_{det} = 1 - \sum_{i=1}^{C_N^0} p_i ^{(0)} -
      \sum_{i=1}^{C_N^1} p_i ^{(1)} -
      ... -
      \sum_{i=1}^{C_N^{n-1}} p_i ^{(n-1)},
\end{equation}
$p_i ^{(0)}$ is the probability of a situation when all antennas are silent,
$p_i ^{(1)}$ --- when one of the antennas detects a signal, but all others do not, and so on.
$i$ is the index of the combinations.
Since the number of the given situations depends on the total number of antennas $N$
and the number of required coincidences $n$, the number of possible combinations is equal to 
the binomial coefficient $C_N^m$. For the case of Tunka-Rex $n$ equals to 3.

The comparison of the footprint-based and LDF-based models is shown in Fig.~\ref{efficiency}.
It can be seen that the latter one provides a smooth distribution of probabilities.
It correctly takes into account the asymmetry of the radio LDF
and the fact that the highest-amplitude signals are not directly
at the shower core position.

\section{Validation of the model}
The Tunka-Rex radio instrument is the only radio cosmic-ray observatory that has
a co-located detector for air-Cherenkov light.
This unique possibility enables validation tests of efficiency models not only with
simulations, but also with the hybrid measurements.
Performing the comparison with a Cherenkov instrument is preferable over a particle instrument.
Both, the Cherenkov- and radio detectors, measure the electromagnetic component of the shower,
while the particle instruments are sensitive to the muonic component as well.
This suppresses the influence of fluctuations due to presence of the
different shower components.

The analyzed data set was specially pre-processed in the framework
of the standard Tunka-Rex analysis
excluding the stages related to the reconstruction
to suppress the influence of the reconstruction efficiency on the results of the validation,
and to study the pure detection efficiency.

\begin{figure}[h!]
  \includegraphics[width=\linewidth]{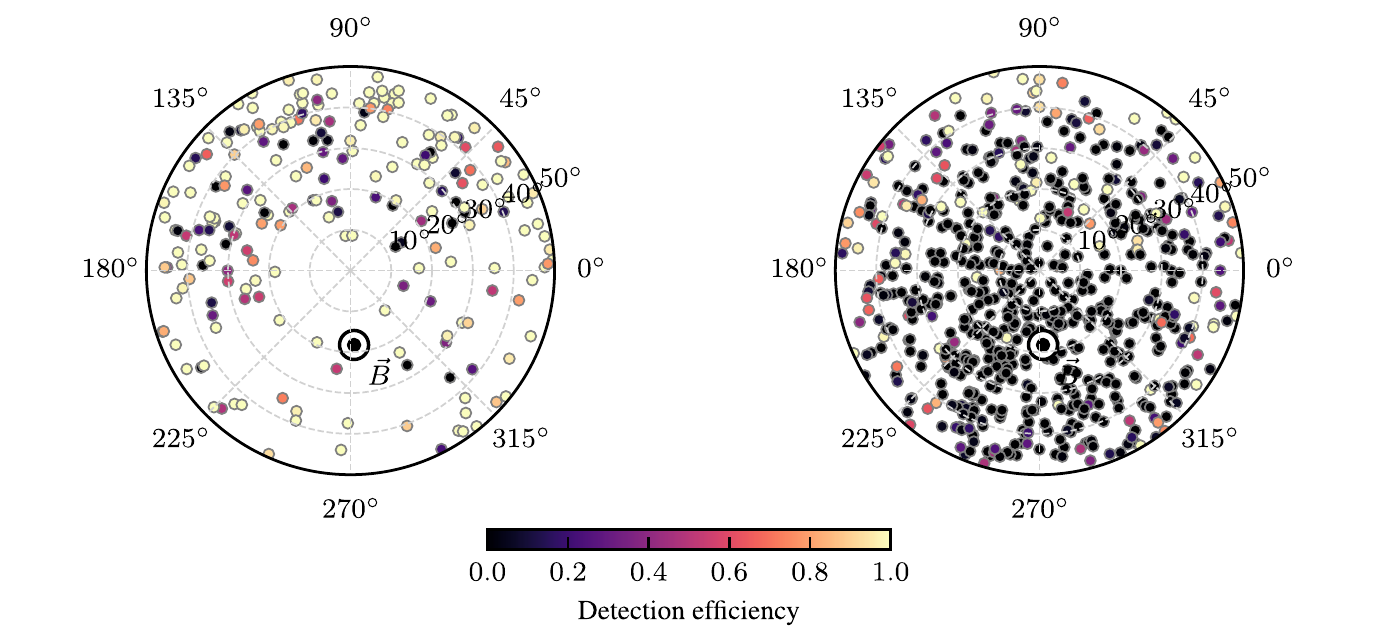}
  \vspace{-8mm}
  \caption{Performance of the new probabilistic LDF-based efficiency model on the measured data. 
Sky maps.
  $\lg$(E/eV)\,>\,17.
  \textit{Left}: the radio events with 3-fold and higher coincidences (events detected by Tunka-133 
and Tunka-Rex).
  \textit{Right}: the radio events with less then 3-fold coincidences (events detected by Tunka-133 
but not seen by Tunka-Rex).}
  \label{comparison}
\end{figure}

The comprehensive validation of the model is ongoing.
The first results of the model performance are shown in Fig.~\ref{comparison}.
It shows the distributions of the data events in the sky and the corresponding detection 
efficiency predicted by the LDF-based model.
One can see the clear efficiency suppression close the direction of the
geomagnetic field, which is expected due to the dominant character of the geomagnetic
mechanism in the radio emission in the traditional frequency band.
This is very well predicted by the model.

\section{Discussion and Outlook}
The estimation of the aperture and efficiency of radio arrays used for cosmic-ray measurements is challenging.
In the present work, we proposed a new approach to the evaluation of these quantities and their 
particular implementation for our instrument.
The new approach includes the information about the antenna detection probability, more realistic spatial signal distribution,
and the probabilistic treatment of the detection process.
These components make the new model conceptually more advanced then the previous one.
The overall accuracy and performance of the model 
and the potential means of the improvement are under investigation.

The present work is an important step towards
a better understanding of the detection efficiency of the radio detectors in general
as well as its different components.
The proper modelling of the detection efficiency provides a way to select 
the spatial regions of the instrument and corresponding cosmic-ray incoming directions 
with the full detection efficiency.
This allows one to reconstruct the cosmic-ray spectrum with lower uncertainties.
Moreover, the well-chosen full-efficiency regions for a radio instrument
open an opportunity to reconstruct the cosmic-ray mass composition with a high accuracy,
in particular when combined with the observations from an instrument of other type.

\section{Acknowledgments}
\footnotesize{
The construction of Tunka-Rex was funded by the German Helmholtz association and the
Russian Foundation for Basic Research (Grant No. HRJRG-303).
This work has been supported
by the Helmholtz Alliance for Astroparticle Physics (HAP), the Russian Federation Ministry
of Science and high Education (Tunka shared core facilities, unique identifier RFMEFI59317X0005,
agreements: 3.9678.2017/8.9, 3.904.2017/4.6),
the Russian Foundation for Basic Research (grant 18-32-00460), 
the grant 19-72-20067 of the Russian Science Foundation (section 5),
by the German Academic Exchange Service (personal grant, ref. num. 91657437).
In preparation of this work we used calculations performed on the HPC-cluster Academician V.\,M.~Matrosov~\cite{HPC_Matrosov}
and on the computational resource ForHLR II funded by the Ministry of Science, Research and the Arts Baden-W\"urttemberg and
DFG (``Deutsche Forschungsgemeinschaft'').
A part of the data analysis was performed using the radio extension of the
Offline framework developed by the Pierre Auger Collaboration~\cite{ARGIRO20071485}.
}

\bibliography{references}
\bibliographystyle{unsrt}

\end{document}